\documentclass[11pt]{article}

\usepackage{pepadun}
\usepackage{caption}
\usepackage{booktabs}   
\usepackage{siunitx}    
\usepackage{tabularx} 
\usepackage{float}
\usepackage{adjustbox}
\usepackage{subcaption}
\usepackage{abstract}
\renewcommand{\abstractnamefont}{\normalfont\bfseries\centering}

\makeatletter
\renewenvironment{abstract}
  {%
    \begin{center}
      {\abstractnamefont\abstractname\par}
    \end{center}
    \vspace{-1.5em}
    \begin{center}
    \small
    \begin{minipage}{0.85\textwidth}
    \setlength{\parindent}{0pt}
  }
  {%
    \end{minipage}
    \end{center}
  }
\makeatother

\newcolumntype{Y}{>{\centering\arraybackslash}X}

\title{Interpreting Hierarchical Composite Endpoints with Survival and Longitudinal Outcomes: Application to Amyotrophic Lateral Sclerosis Trials}
\author{%
Arlina Shen\textsuperscript{1}, 
Dehua Bi\textsuperscript{1},  
Ruben P.A. van Eijk\textsuperscript{2}, 
Lu Tian\textsuperscript{1},
and Ying Lu\textsuperscript{1} \\[6pt]
\textsuperscript{1} Department of Biomedical Data Science, School of Medicine, Stanford University, Stanford, CA 94305, USA \\
\textsuperscript{2} Department of Neurology, UMC Utrecht Brain Center, University Medical Center Utrecht, Utrecht, The Netherlands
}
\date{\today} 

\begin{document}

\maketitle
\pagestyle{plain}
\begin{abstract}

{\color{black}Hierarchical composite endpoints combining survival and longitudinal functional outcomes are increasingly used in clinical trials, especially when death precludes subsequent functional assessment. The Finkelstein--Schoenfeld strategy analyzes such endpoints through prioritized pairwise comparisons, with survival compared before function. In amyotrophic lateral sclerosis (ALS), this strategy is implemented in the Combined Assessment of Function and Survival (CAFS), which combines survival with the ALS Functional Rating Scale Revised. Although such endpoints provide a clinically meaningful summary of overall treatment benefit, investigators may also want to understand whether the treatment effect is driven by survival, functional outcome, or both. Motivated by the estimand framework, we use ALS-informed simulations from a joint longitudinal--survival model to study settings in which treatment effects on survival and function align or conflict. The simulations show that decomposing the composite win probability into survival and survivor-based functional contributions clarifies their relative roles and separates the composite treatment-benefit question from function-focused questions, including the while-alive comparison and conceptual alternatives based on hypothetical and always-survivor estimands. The functional win probability underlying the while-alive comparison can be subject to survivor-selection bias when treatment affects survival, and inverse-probability weighting can attenuate selection induced by measured predictors under appropriate assumptions. Brief supporting analyses based on principal stratification and multiply robust estimation illustrate the always-survivor estimand. This framework provides practical guidance for reporting and interpreting hierarchical composite endpoints, with CAFS in ALS serving as a concrete motivating example.}


\end{abstract}


\section{Introduction}

Clinical trials in progressive diseases often collect both time-to-event outcomes, such as death or major clinical intervention, and longitudinal outcomes, such as functional status or disease progression measured repeatedly over follow-up. Hierarchical composite endpoints combine such outcomes by ordering clinical events according to priority or desirability and then summarizing the resulting ranks or pairwise wins. The Finkelstein--Schoenfeld framework was introduced to combine mortality and longitudinal outcomes through prioritized pairwise comparisons: treated--control pairs are compared first on the highest-priority outcome, often survival, and then on lower-priority outcomes for pairs not distinguished by the higher-priority outcome \cite{FinkelsteinSchoenfeld1999}. The method belongs to the broader class of generalized pairwise comparisons (GPC), which also includes win-ratio and related win-statistic procedures \cite{Buyse2010GPC,Verbeeck2019GPC,ParnerOvergaard2023,Cui2025}. A closely related rank-based approach is desirability of outcome ranking (DOOR), which compares treated--control pairs on a prespecified ordinal outcome and estimates the probability of a more desirable outcome \cite{evans2015door}. Percentile-based composite rankings, such as the analysis used in the N-TA3CT abdominal aortic aneurysm (AAA) trial, provide another related rank-based approach \cite{Baxter2020}.

These methods combine survival and longitudinal outcomes according to prespecified priorities and are particularly relevant when a survival endpoint alone does not capture clinically meaningful longitudinal information and a longitudinal analysis is complicated by death. The Finkelstein--Schoenfeld ranking therefore defines a composite estimand that is reported as a single summary of overall treatment benefit. In practice, investigators may also want to understand treatment effects on its individual outcomes, which the overall win probability does not directly reveal. Such interpretation is especially challenging for a lower-priority endpoint in a hierarchical GPC because its comparison is made only among pairs not distinguished by higher-priority outcomes. In a survival-first composite, for example, the functional outcome is compared only for pairs in which both participants are alive at the relevant follow-up time. When treatment affects survival, the survivor-conditional functional comparison can be subject to survivor-selection bias if interpreted as a function-specific treatment effect in a common target population. This study therefore decomposes the composite to characterize its survival and functional contributions, examine how differential survival affects the lower-priority functional comparison, and determine whether adjustment for measured predictors can attenuate this selection across settings in which survival and functional effects align or conflict.

We use amyotrophic lateral sclerosis (ALS) as a motivating example because both survival and functional decline are clinically central, and the functional outcome is truncated by death. In ALS trials, the Finkelstein--Schoenfeld strategy is implemented as the Combined Assessment of Function and Survival (CAFS), a disease-specific hierarchical composite that uses survival and the 12-item ALS Functional Rating Scale Revised (ALSFRS-R) to rank treated--control pairs \cite{evans2016using,Meininger2017ozanezumab,berry2013cafs}. {\color{black}FDA guidance for ALS drug development similarly recommends integrating survival and functional outcomes into a single overall measure when the functional outcome is primary \cite{FDA2019ALS}.}

Using ALS-motivated simulations and counterfactual analyses under a joint longitudinal--survival model \cite{vanEijk2022joint}, we study three main settings: survival benefit without functional benefit, survival harm with functional benefit, and survival and functional benefit. We decompose the CAFS win probability into survival and survivor-based functional contributions and evaluate inverse-probability-weighted adjustment of the functional comparison. Supporting principal-stratum analyses and multiply robust estimation illustrate the distinct always-survivor target. Our contribution is an estimand-focused clarification of how hierarchical composite endpoints behave when survival and functional effects align or conflict and how survivor-selection bias can affect the survivor-conditional functional comparison. We further distinguish IPW adjustment of this comparison from principal-stratum analysis of the always-survivor estimand and relate both to the composite target. Section 2 introduces the notation, observed-data estimands, the composite decomposition, and the joint model. Section 3 presents the simulation results and discusses their implications for interpreting the composite and its component-specific estimands, with supporting diagnostic and principal-stratum analyses. Section 4 concludes with practical implications for interpreting hierarchical composite analyses.

\section{Methods}

\subsection{Basic Setup and Notation} 
Consider a randomized two-arm clinical trial with $n_1$ participants assigned to treatment and $n_0$ participants assigned to control. Let $A_i \in \{0,1\}$ denote the treatment assignment for participant $i$, with $A_i=1$ for treatment and $A_i=0$ for control. Let $X_i$ denote baseline covariates. For each participant, let $T_i$ denote survival time from baseline and let $Y_i(t)$ denote the functional outcome measured at time $t$ (ALSFRS-R), when defined. 

Because death is an intercurrent event for the functional outcome, 
$Y_i(t)$ is only defined 
when $T_i >t$. This distinction leads naturally to different estimands depending on whether the primary scientific question concerns survival, functional status at a fixed time, or a composite ranking of both outcomes. In particular, when treatment affects survival, a comparison based only on observed functional outcomes among survivors need not match either a survival-based comparison or a principal-stratum contrast among participants who would survive under either treatment.

For population-level quantities, let $\{T^{(1)},Y^{(1)}(t)\}$ and $\{T^{(0)},Y^{(0)}(t)\}$ denote generic outcome pairs from the treatment and control arms, respectively, where $Y^{(a)}(t)$ is defined only when $T^{(a)}>t$, for $a\in\{0,1\}$. Let
\[
S_a(t)=P(T^{(a)}>t), \qquad a\in\{0,1\},
\]
denote the corresponding survival functions.


\subsection{Pairwise Win Indicators for Survival and Functional Outcomes}

{\color{black}Rank-based pairwise comparisons evaluate a treated participant against a control participant and assign a win, loss, or tie according to their observed outcomes.} Let treated participants be indexed by $i=1,\dots,n_1$ and control participants by $j=1,\dots,n_0$. 
For survival at time $t$, define the pairwise survival 
win indicator
\[
W_{ij,S}(t)=
\begin{cases}
1, & T_i > T_j \ \text{and}\ T_j \le t,\\
0, & T_i < T_j \ \text{and}\ T_i \le t,\\
0.5, & T_i > t\ \text{and}\ T_j > t,\\
0.5, & T_i = T_j \le t.
\end{cases}
\]
Under a continuous-time survival model, $P(T_i=T_j\le t)=0$, so the only systematic tie arises 
when both participants survive beyond $t$.

The corresponding estimator is the 
U-statistic 
\[
\widehat\theta_{S}(t)
=\frac{1}{n_1n_0}\sum_{i=1}^{n_1}\sum_{j=1}^{n_0}W_{ij,S}(t),
\]
which estimates the population survival win probability 
\[
\theta_S(t)
= P\left(T^{(1)} > T^{(0)}, \, T^{(0)} \le t\right)
+ \tfrac{1}{2} P(T^{(1)} > t, \, T^{(0)} > t). \]
Equivalently,
\[
\theta_S(t) = -\int_{0}^{t}S_{1}(u)\,\mathrm{d}S_{0}(u)
\;+\;
\tfrac{1}{2}\,S_{1}(t)\,S_{0}(t).
\] 
Values of $\theta_S(t)$ above 
$0.5$ favor treatment, whereas values 
below $0.5$ favor control.

For the functional outcome, comparisons at time $t$ are only defined for pairs in which both participants survive beyond $t$. For such pairs, define the functional win indicator
\[
W_{ij,Y}(t) =
\begin{cases}
1,   & Y_i(t)>Y_j(t),\\
0,   & Y_i(t)<Y_j(t),\\
0.5, & Y_i(t)=Y_j(t).
\end{cases}
\] 
The corresponding survivor-conditional functional win probability is 
\[
\theta_Y(t)
=
P\!\bigl\{Y^{(1)}(t)>Y^{(0)}(t)\mid T^{(1)}>t,\,T^{(0)}>t\bigr\}
+\tfrac{1}{2}P\!\bigl\{Y^{(1)}(t)=Y^{(0)}(t)\mid T^{(1)}>t,\,T^{(0)}>t\bigr\}.
\] 
{\color{black}We refer to $\theta_Y(t)$ as the survivor-conditional functional win probability. It supplies the functional comparison used to form the survivor-based contribution in the CAFS decomposition below.} Without further assumptions, $\theta_Y(t)$ is not a marginal causal effect on function or an always-survivor causal contrast. IPW-adjusted and principal-stratum quantities introduced below target distinct functional comparisons rather than alternative estimators of this same population quantity.

\subsection{CAFS as a Hierarchical Composite Estimand}

As the ALS-specific implementation considered here, CAFS combines survival and functional outcome through survival-first pairwise comparisons. Specifically, survival determines the pairwise ordering whenever the treated and control participants can be distinguished by survival up to time $t$; 
only for pairs in which 
both participants 
survive beyond $t$.

Accordingly, define the CAFS pairwise win indicator for treated participant $i$ and control participant $j$ by
\[
W_{ij,\mathrm{CAFS}}(t)=
\begin{cases}
1, & T_i>T_j \text{ and }\ T_j\le t,\\
0, & T_i<T_j \text{ and }\ T_i\le t,\\
W_{ij,Y}(t), & T_i>t \text{ and }\ T_j>t,\\
0.5, & T_i=T_j\le t.
\end{cases}
\]
Under a continuous-time survival model, the event $T_i=T_j\le t$ has probability zero, so the last line is included only for completeness.

The corresponding population CAFS win probability is
\[
\theta_{\mathrm{CAFS}}(t)=E[W_{ij,\mathrm{CAFS}}(t)].
\]
Using the generic arm-specific notation introduced above, this expectation may be written as 
\[
\theta_{\mathrm{CAFS}}(t)
=
P(T^{(1)}>T^{(0)},\,T^{(0)}\le t)
+
E\!\left[
W_Y(t)\mid T^{(1)}>t,\,T^{(0)}>t
\right]
P(T^{(1)}>t,\,T^{(0)}>t),
\]
where $W_Y(t)$ denotes the functional comparison for a generic treated--control pair among participants who both survive beyond $t$. By definition,
\[
E\!\left[
W_Y(t)\mid T^{(1)}>t,\,T^{(0)}>t
\right]
=
\theta_Y(t),
\]
so that
\[
\theta_{\mathrm{CAFS}}(t)
=
P(T^{(1)}>T^{(0)},\,T^{(0)}\le t)
+
\theta_Y(t)\,P(T^{(1)}>t,\,T^{(0)}>t).
\]
Under independence of the two treatment arms,
\[
P(T^{(1)}>t,\,T^{(0)}>t)=S_1(t)S_0(t),
\]
and the probability that the treated participant wins on survival before time $t$ is
\[
P(T^{(1)}>T^{(0)},\,T^{(0)}\le t)
=
-\int_0^t S_1(u)\,dS_0(u).
\]
Substituting these expressions into the previous display yields

\begin{equation}\label{eq:theta_cafs}
    \theta_{\mathrm{CAFS}}(t)
=
-\int_{0}^{t}S_{1}(u)\,\mathrm{d}S_{0}(u)
\;+\;
\theta_{Y}(t)\,S_{1}(t)\,S_{0}(t)
=
\theta_{S}(t)\;+\;
\left[\theta_{Y}(t)\,-0.5\,\right]\,S_{1}(t)\,S_{0}(t)
.
\end{equation}
{\color{black}This representation makes explicit how the composite ranking depends on both contributions. Survival enters through the first contribution, whereas function contributes only among pairs in which both participants survive beyond $t$. This distinction is central to our simulations, in which survival and functional effects may align or conflict, because it clarifies how the two contributions jointly determine the overall CAFS result while addressing different component-specific questions.}

\subsection{Inverse-Probability Weighting for Survivor-Conditional Functional Comparisons}
{\color{black}In observed data, the functional outcome at time $t$ is available only for participants who survive beyond $t$. The observed-survivor comparison is a well-defined while-alive functional summary, but treatment-related differences in survival can change the covariate composition of survivors in each arm. We therefore use inverse-probability-of-survival weighting to reweight this comparison according to participants' probabilities of remaining alive at $t$. The resulting target is an IPW-adjusted while-alive functional comparison, distinct from both the unadjusted survivor-conditional quantity and the always-survivor estimand.}

Let $R_i(t)=I\{T_i>t\}$ indicate whether participant $i$ survives beyond $t$ and therefore contributes to the functional comparison. Define
\[
G_a(x,t)=P(T_i>t\mid A_i=a,X_i=x), \qquad a\in\{0,1\},
\]
{\color{black}Using the arm-specific notation introduced above, the population IPW-adjusted while-alive functional comparison is}
\[
\theta_Y^{\mathrm{IPW}}(t)
=
\frac{
E\!\left[
\dfrac{I\{T^{(1)}>t\}I\{T^{(0)}>t\}W_Y(t)}
   {G_1(X^{(1)},t)G_0(X^{(0)},t)}
\right]
}{
E\!\left[\dfrac{I\{T^{(1)}>t\}}{G_1(X^{(1)},t)}\right]
E\!\left[\dfrac{I\{T^{(0)}>t\}}{G_0(X^{(0)},t)}\right]
}.
\]
{\color{black}This target compares observed functional outcomes after reweighting survivors toward the baseline-covariate distributions of their randomized arms. The corresponding IPW estimator is}
\[
\widehat\theta_{Y}^{\mathrm{IPW}}(t)
=
\frac{
  \displaystyle
  \sum_{i=1}^{n_1}\sum_{j=1}^{n_0}
    \frac{R_i(t)R_j(t)\,W_{ij,Y}(t)}
         {G_{1i}(t)\,G_{0j}(t)}
}{
  \displaystyle
  \left[
    \sum_{i=1}^{n_1}\frac{R_i(t)}{G_{1i}(t)}
  \right]
  \left[
    \sum_{j=1}^{n_0}\frac{R_j(t)}{G_{0j}(t)}
  \right]
}.
\]
{\color{black}Here, $G_{1i}(t)=G_1(X_i,t)$ and $G_{0j}(t)=G_0(X_j,t)$.}
{\color{black}In applications, $G_{ai}(t)$ can be estimated from an appropriate survival model using observed prognostic covariates.}
{\color{black}For illustration, replacing $\widehat\theta_Y(t)$ in the sample CAFS decomposition with the reweighted functional comparison gives the constructed estimator}
\[
\widehat\theta_{\mathrm{CAFS}}^{\mathrm{IPW}}(t)
=
-\int_0^t \widehat S_1(u)\,d\widehat S_0(u)
+
\widehat\theta_Y^{\mathrm{IPW}}(t)\,\widehat S_1(t)\widehat S_0(t).
\]
{\color{black}This constructed IPW-adjusted CAFS-type summary combines the original survival contribution with a reweighted functional contribution and is therefore distinct from the original CAFS estimand in~\eqref{eq:theta_cafs}. For brevity, we refer to it as IPW-CAFS. Under positivity and consistently estimated survival probabilities, weighting balances the measured covariates included in $X$. Interpreting this balance as attenuation of survivor selection requires $X$ to include the relevant measured predictors of survival and observed function. The functional target remains a while-alive comparison and does not identify function after death or the always-survivor estimand.}

\subsection{Joint Longitudinal--Survival Model}
To capture the dependence between the functional trajectory and survival, we adopt the joint-model specification of van Eijk \emph{et al.} \cite{vanEijk2022joint}. Let $A_i\in\{0,1\}$ denote treatment assignment. {\color{black}The latent longitudinal process is
\begin{align}
\eta_i(t) &= B_{0i}+B_{1i}t+B_{2i}t^2+B_4 A_i t, \\
Y_i(t) &= \eta_i(t)+\varepsilon_{it}, \qquad \varepsilon_{it}\stackrel{i.i.d.}{\sim}N(0,\sigma^2),
\end{align}
with participant-specific random effects
\[
\begin{pmatrix}B_{0i}\\ B_{1i}\\ B_{2i}\end{pmatrix}
\sim
N\!\left(
\begin{pmatrix}B_0\\ B_1\\ B_2\end{pmatrix},
\mathbf{\Sigma}
\right),
\qquad
(B_{0i},B_{1i},B_{2i}) \perp \{\varepsilon_{it}:t\ge0\}.
\]}
The survival submodel uses a Weibull baseline hazard with time-dependent association to the latent functional trajectory, 
\begin{equation}
h_i(t\mid B_{0i},B_{1i},B_{2i})
= \rho\, t^{\rho-1}\,
\exp\!\left\{
g_0 + g_1 A_i + \alpha_1 \eta_i(t) + \alpha_2 \eta_i'(t)
\right\},
\label{eq:weibull-haz}
\end{equation}
{\color{black}where $\eta_i'(t)=B_{1i}+2B_{2i}t+B_4A_i$.} The corresponding conditional survival function is \begin{equation}
S_i(t\mid B_{0i},B_{1i},B_{2i})
=
\exp\!\left\{
-\int_{0}^{t}
\rho\,u^{\rho-1}\,
\exp\!\left[g_0 + g_1 A_i + \alpha_1 \eta_i(u) + \alpha_2 \eta_i'(u)\right]
\,du
\right\}.
\label{eq:weibull-surv}
\end{equation} {\color{black}In the longitudinal submodel, $B_4$ represents the treatment-associated change in the functional trajectory. In the survival submodel, $g_1$ represents the direct treatment effect on the survival hazard, whereas $\alpha_1$ and $\alpha_2$ quantify} 
the association of survival with the latent current level ($\eta_i(t)$) and slope ($\eta_i'(t)$), respectively.

{\color{black}The joint model of van Eijk \emph{et al.} \cite{vanEijk2022joint} is a parametric likelihood-based framework: a longitudinal mixed model for the latent functional trajectory is coupled with a fully specified time-to-event model, with association parameters linking event risk to features of the latent trajectory. This structure makes explicit which model components are relevant for different estimands. Neither the original CAFS estimator nor the constructed IPW-adjusted CAFS-type estimator requires specifying a full joint likelihood for $(Y(t),T)$. Rather than relying on such a likelihood, CAFS is computed directly from observed prioritized pairwise comparisons. The IPW-adjusted analysis uses conditional survival probabilities, which may be estimated from a separate survival model, to reweight the survivor-conditional functional comparison and attenuate selection explained by measured prognostic covariates.}

{\color{black}For the simulation study, we nevertheless generate data from the parametric joint model because it provides a transparent and controllable mechanism in which the direct treatment parameters for survival and the functional trajectory can be varied separately, while the survival--function dependence that induces survivor-selection bias is explicit. The model is particularly suitable for this purpose because it was informed by the real randomized valproic acid (VPA) trial in ALS, giving a clinically grounded yet interpretable data-generating process. This allows us to assess IPW when the source of informative survival is known and scientifically plausible, while keeping the focus on the estimands and the adjustment role of IPW rather than on committing to a specific parametric analysis model.}
\section{Numerical Results and Discussion}

\subsection{Simulation Design}
We conducted a Monte Carlo study under the joint longitudinal--survival data-generating mechanism described in the Methods section, 
adapted from the framework that van Eijk \emph{et al.} applied to the randomized VPA trial in ALS \cite{vanEijk2022joint}. {\color{black}We simulated survival times and longitudinal ALSFRS-R scores and evaluated $\widehat\theta_S(t)$, $\widehat\theta_Y(t)$, $\widehat\theta_Y^{\mathrm{IPW}}(t)$, $\widehat\theta_{\mathrm{CAFS}}(t)$, and the constructed IPW-adjusted CAFS-type summary $\widehat\theta_{\mathrm{CAFS}}^{\mathrm{IPW}}(t)$. The simulations are intended to characterize estimand behavior and selection patterns. The three primary settings represent survival benefit without functional benefit, survival harm with functional benefit, and benefit in both domains; null and survival-harm-only scenarios serve as reference cases.} Each primary Monte Carlo experiment used 1,000 independent replications with $N=500$ participants per replication. {\color{black}No censoring, missed visits, or dropout was imposed, allowing the comparisons to isolate selection associated with death. Hatted entries in the tables are Monte Carlo means of the sample estimators across replications. Table~\ref{tab:dgp-common} summarizes the common simulation parameters. Observed-data results are presented with selected counterfactual reference values. Section~\ref{subsec:counterfactual-ps} introduces the arm-specific parameter choices for the counterfactual framework and examines the principal-stratum estimands.}

\begin{table}[H]
\centering
\caption{\small {\color{black}Parameters common to the joint-model simulations informed by the randomized VPA trial in ALS analyzed by van Eijk \emph{et al.} \cite{vanEijk2022joint}. The treatment parameters $B_4$ and $g_1$ vary by scenario.}}
\label{tab:dgp-common}
\small
\begingroup
\setlength{\tabcolsep}{5pt}
\renewcommand{\arraystretch}{1.14}
\begin{tabular}{@{}>{\raggedright\arraybackslash}m{0.43\textwidth} >{\centering\arraybackslash}m{0.10\textwidth} >{\centering\arraybackslash}m{0.37\textwidth}@{}}
\toprule
Parameter & Symbol & Value \\
\midrule
{\color{black}Mean baseline ALSFRS-R score} & {\color{black}$B_0$} & {\color{black}$39.242$} \\
{\color{black}Mean annual ALSFRS-R slope under control} & {\color{black}$B_1$} & {\color{black}$-14.329$} \\
{\color{black}Mean quadratic ALSFRS-R coefficient under control} & {\color{black}$B_2$} & {\color{black}$1.306$} \\
Residual variance of ALSFRS-R measurement error & {\color{black}$\sigma^2$} & {\color{black}$2.965$} \\
Weibull hazard shape parameter & {\color{black}$\rho$} & {\color{black}$1.050$} \\
Baseline log-hazard intercept & {\color{black}$g_0$} & {\color{black}$0.245$} \\
Log-hazard association with current ALSFRS-R score & {\color{black}$\alpha_1$} & {\color{black}$-0.088$} \\
Log-hazard association with ALSFRS-R slope & {\color{black}$\alpha_2$} & {\color{black}$-0.049$} \\
{\color{black}Random-effects covariance for participant-specific baseline score, slope, and quadratic coefficient} &
{\color{black}$\mathbf{\Sigma}$} &
{\color{black}$\displaystyle
\begin{bmatrix}
36.764 & 28.104 & -14.707 \\
28.104 & 177.171 & -58.264 \\
-14.707 & -58.264 & 28.654
\end{bmatrix}$} \\
\bottomrule
\end{tabular}
\endgroup
\end{table}

\subsection{Setting 1: Survival Benefit Without Functional Benefit}

{\color{black}Because the simulation provides potential outcomes under both treatments, Table~\ref{tab:mc-results} includes two counterfactual benchmarks. The functional win probability $\theta_Y^{\mathrm{AS}}(t)$ is defined among participants who would survive beyond $t$ under either treatment and is a principal-stratum estimand. The constructed benchmark $\theta_{\mathrm{CAFS}}^{\mathrm{AS}}(t)$ replaces $\theta_Y(t)$ in the survivor-based functional contribution with $\theta_Y^{\mathrm{AS}}(t)$; it is not a conventional principal-stratum estimand. Both quantities are used only as simulation references. Section~\ref{subsec:counterfactual-ps} develops the principal-stratum analysis, and Appendix~\ref{subsec:principal-stratum-assumptions} summarizes its identification assumptions.}

\begin{table}[H]
\centering
\captionsetup{font=footnotesize}
\caption{{\color{black}Monte Carlo results for the null benchmark and Setting~1, including counterfactual always-survivor reference values available from the simulation. Time $t$ is measured in years.}}
\label{tab:mc-results}

\scriptsize
\setlength{\tabcolsep}{4pt}
\renewcommand{\arraystretch}{0.86}

\begin{adjustbox}{max width=\textwidth}
\begin{tabular}{c c c c c c c c c c c}
\toprule
$t$ & $g_1$ & ${\color{black}B_4}$
& $\theta_S(t)$ & $\widehat\theta_S(t)$
& $\theta_Y^{\mathrm{AS}}(t)$ & $\widehat\theta_Y(t)$
& $\widehat\theta_Y^{\mathrm{IPW}}(t)$
& $\theta_{\mathrm{CAFS}}^{\mathrm{AS}}(t)$ & $\widehat\theta_{\mathrm{CAFS}}(t)$
& $\widehat\theta_{\mathrm{CAFS}}^{\mathrm{IPW}}(t)$ \\
\midrule
0.2 & 0.000 & 0.000 & 0.500 & 0.500 & 0.500 & 0.499 & 0.499 & 0.500 & 0.499 & 0.499 \\
0.5 & 0.000 & 0.000 & 0.500 & 0.500 & 0.500 & 0.498 & 0.498 & 0.500 & 0.498 & 0.498 \\
0.8 & 0.000 & 0.000 & 0.500 & 0.500 & 0.500 & 0.498 & 0.498 & 0.500 & 0.498 & 0.498 \\
1.0 & 0.000 & 0.000 & 0.500 & 0.500 & 0.500 & 0.498 & 0.498 & 0.500 & 0.498 & 0.498 \\
1.3 & 0.000 & 0.000 & 0.500 & 0.500 & 0.500 & 0.499 & 0.499 & 0.500 & 0.499 & 0.499 \\
1.6 & 0.000 & 0.000 & 0.500 & 0.500 & 0.500 & 0.500 & 0.500 & 0.500 & 0.500 & 0.499 \\
\midrule
0.2 & $\log(0.5)$ & 0.000 & 0.507 & 0.507 & 0.500 & 0.495 & 0.499 & 0.507 & 0.502 & 0.506 \\
0.5 & $\log(0.5)$ & 0.000 & 0.520 & 0.520 & 0.500 & 0.488 & 0.500 & 0.520 & 0.510 & 0.521 \\
0.8 & $\log(0.5)$ & 0.000 & 0.535 & 0.535 & 0.500 & 0.478 & 0.500 & 0.535 & 0.518 & 0.535 \\
1.0 & $\log(0.5)$ & 0.000 & 0.544 & 0.544 & 0.500 & 0.472 & 0.499 & 0.544 & 0.525 & 0.544 \\
1.3 & $\log(0.5)$ & 0.000 & 0.557 & 0.557 & 0.500 & 0.462 & 0.498 & 0.557 & 0.536 & 0.556 \\
1.6 & $\log(0.5)$ & 0.000 & 0.569 & 0.569 & 0.500 & 0.451 & 0.493 & 0.569 & 0.546 & 0.566 \\
\bottomrule
\end{tabular}
\end{adjustbox}
\end{table}

{\color{black}The null rows ($g_1=0$, $B_4=0$) provide the reference case, with all win probabilities remaining near $0.5$. In Setting~1, we set the treatment effect on the functional trajectory to zero ($B_4=0$) but imposed a survival benefit for treatment ($g_1=\log(0.5)$), so that more treated participants remain alive at later times. Both $\widehat\theta_{\mathrm{CAFS}}(t)$ and the constructed IPW-adjusted CAFS-type summary $\widehat\theta_{\mathrm{CAFS}}^{\mathrm{IPW}}(t)$ increase over time, reflecting the growing survival advantage, but a visible gap emerges between them. This separation is informative because $\widehat\theta_Y(t)$ falls below $0.5$, reaching $0.451$ at $t=1.6$, even though the direct functional treatment effect is null and $\theta_Y^{\mathrm{AS}}(t)$ remains $0.5$. The decline in $\widehat\theta_Y(t)$ is therefore not evidence of a harmful direct functional treatment effect; rather, it reflects survivor-selection bias when the observed survivor comparison is interpreted as a functional effect in a common target population.}

{\color{black}The role of IPW is to attenuate selection induced by measured predictors in the survivor-conditional functional comparison under the weighting assumptions. In that sense, $\theta_{\mathrm{CAFS}}^{\mathrm{IPW}}(t)$ is best viewed as an adjusted CAFS-type summary, not as the same estimand as the original CAFS win probability. Here, the IPW-CAFS summary stays more closely aligned with the survival-driven part of the decomposition, showing that the difference between original CAFS and IPW-CAFS arises from adjustment of the survivor-conditional functional comparison while both summaries remain within a survival-first composite framework. Results using the generated random effects showed a similar small late-follow-up deviation as the effective sample size declined, consistent with finite-sample instability under increasingly selective survival rather than misspecified survival probabilities.}

\subsection{Setting 2: Survival Harm With Functional Benefit}

\begin{table}[H]
\centering
\captionsetup{font=footnotesize}
\caption{
{\color{black}Monte Carlo estimates at time $t$ (in years) for the survival-harm setting. The first block is a survival-harm-only reference scenario ($g_1=\log(2)$, ${\color{black}B_4}=0$). The second block is Setting~2, in which treatment harms survival ($g_1=\log(2)$) but improves the functional trajectory (${\color{black}B_4}=2$). Columns follow the same notation as in Table~\ref{tab:mc-results}.}
}
\label{tab:mc-results-harmful}

\scriptsize
\setlength{\tabcolsep}{4pt}
\renewcommand{\arraystretch}{0.86}

\begin{adjustbox}{max width=\textwidth}
\begin{tabular}{c c c c c c c c c c c}
\toprule
$t$ & $g_1$ & ${\color{black}B_4}$
& $\theta_S(t)$ & $\widehat\theta_S(t)$
& $\theta_Y^{\mathrm{AS}}(t)$ & $\widehat\theta_Y(t)$
& $\widehat\theta_Y^{\mathrm{IPW}}(t)$
& $\theta_{\mathrm{CAFS}}^{\mathrm{AS}}(t)$ & $\widehat\theta_{\mathrm{CAFS}}(t)$
& $\widehat\theta_{\mathrm{CAFS}}^{\mathrm{IPW}}(t)$ \\
\midrule
0.2 & $\log(2)$ & 0.000 & 0.487 & 0.487 & 0.500 & 0.508 & 0.501 & 0.487 & 0.494 & 0.488 \\
0.5 & $\log(2)$ & 0.000 & 0.466 & 0.465 & 0.500 & 0.520 & 0.499 & 0.466 & 0.481 & 0.465 \\
0.8 & $\log(2)$ & 0.000 & 0.446 & 0.446 & 0.500 & 0.534 & 0.500 & 0.446 & 0.467 & 0.446 \\
1.0 & $\log(2)$ & 0.000 & 0.436 & 0.435 & 0.500 & 0.543 & 0.502 & 0.436 & 0.458 & 0.437 \\
1.3 & $\log(2)$ & 0.000 & 0.423 & 0.423 & 0.500 & 0.556 & 0.513 & 0.423 & 0.445 & 0.428 \\
1.6 & $\log(2)$ & 0.000 & 0.413 & 0.413 & 0.500 & 0.568 & 0.520 & 0.413 & 0.433 & 0.419 \\
\midrule
0.2 & $\log(2)$ & 2.000 & 0.490 & 0.490 & 0.514 & 0.524 & 0.519 & 0.503 & 0.512 & 0.507 \\
0.5 & $\log(2)$ & 2.000 & 0.474 & 0.474 & 0.531 & 0.546 & 0.530 & 0.499 & 0.510 & 0.497 \\
0.8 & $\log(2)$ & 2.000 & 0.462 & 0.461 & 0.543 & 0.565 & 0.537 & 0.490 & 0.503 & 0.485 \\
1.0 & $\log(2)$ & 2.000 & 0.455 & 0.455 & 0.552 & 0.580 & 0.548 & 0.484 & 0.500 & 0.482 \\
1.3 & $\log(2)$ & 2.000 & 0.449 & 0.449 & 0.564 & 0.596 & 0.559 & 0.477 & 0.491 & 0.474 \\
1.6 & $\log(2)$ & 2.000 & 0.445 & 0.445 & 0.576 & 0.610 & 0.571 & 0.470 & 0.481 & 0.468 \\
\bottomrule
\end{tabular}
\end{adjustbox}
\end{table}
The first block of Table~\ref{tab:mc-results-harmful} is included as a reference scenario showing survival harm without functional benefit. Here, $\theta_S(t)$ falls below $0.5$ and $\theta_Y^{\mathrm{AS}}(t)$ remains $0.5$, so the composite summaries decrease over time and favor control. At the same time, $\widehat\theta_Y(t)$ increases over follow-up, illustrating how a survivor-conditional functional comparison can be distorted when treatment changes who remains alive to be observed.

{\color{black}Setting~2 is a clinically important case in which treatment improves the functional trajectory but is associated with worse survival. In the second block of Table~\ref{tab:mc-results-harmful}, both $\widehat\theta_{\mathrm{CAFS}}(t)$ and $\widehat\theta_{\mathrm{CAFS}}^{\mathrm{IPW}}(t)$ decline over time. The original CAFS estimate begins above $0.5$ because the functional benefit initially offsets the survival disadvantage in the composite ranking, whereas the IPW-adjusted CAFS-type summary falls below $0.5$ by $t=0.5$. At the same time, $\theta_Y^{\mathrm{AS}}(t)$, $\widehat\theta_Y(t)$, and $\widehat\theta_Y^{\mathrm{IPW}}(t)$ remain above $0.5$, so the functional comparisons continue to favor treatment even while $\theta_S(t)$ remains below $0.5$. The observed-survivor comparison $\widehat\theta_Y(t)$ increases more strongly over follow-up, but this trend requires caution because treatment changes the survivor population represented at each time. The more muted trend in $\widehat\theta_Y^{\mathrm{IPW}}(t)$ is consistent with attenuation of survivor-selection bias associated with the measured predictors used in the weights.}

{\color{black}The later time points make the survival-first nature of the hierarchical composite especially clear. As follow-up increases, the harmful survival effect contributes more strongly to the pairwise ranking, so the composite is pulled downward even though the functional comparisons remain favorable to treatment. Thus, IPW does not make the composite more favorable to treatment in this setting; rather, it shows that after attenuating survivor-selection bias in the survivor-conditional functional comparison, the accumulating survival harm dominates the composite comparison.}
\subsection{Setting 3: Survival and Functional Benefit}

Finally, we consider Setting~3, in which treatment confers both a survival benefit, $g_1 = \log(0.5)$, and a direct functional benefit, ${\color{black}B_4} = 2$.

\begin{figure}[H]
  \centering
  \includegraphics[width=0.7\textwidth]{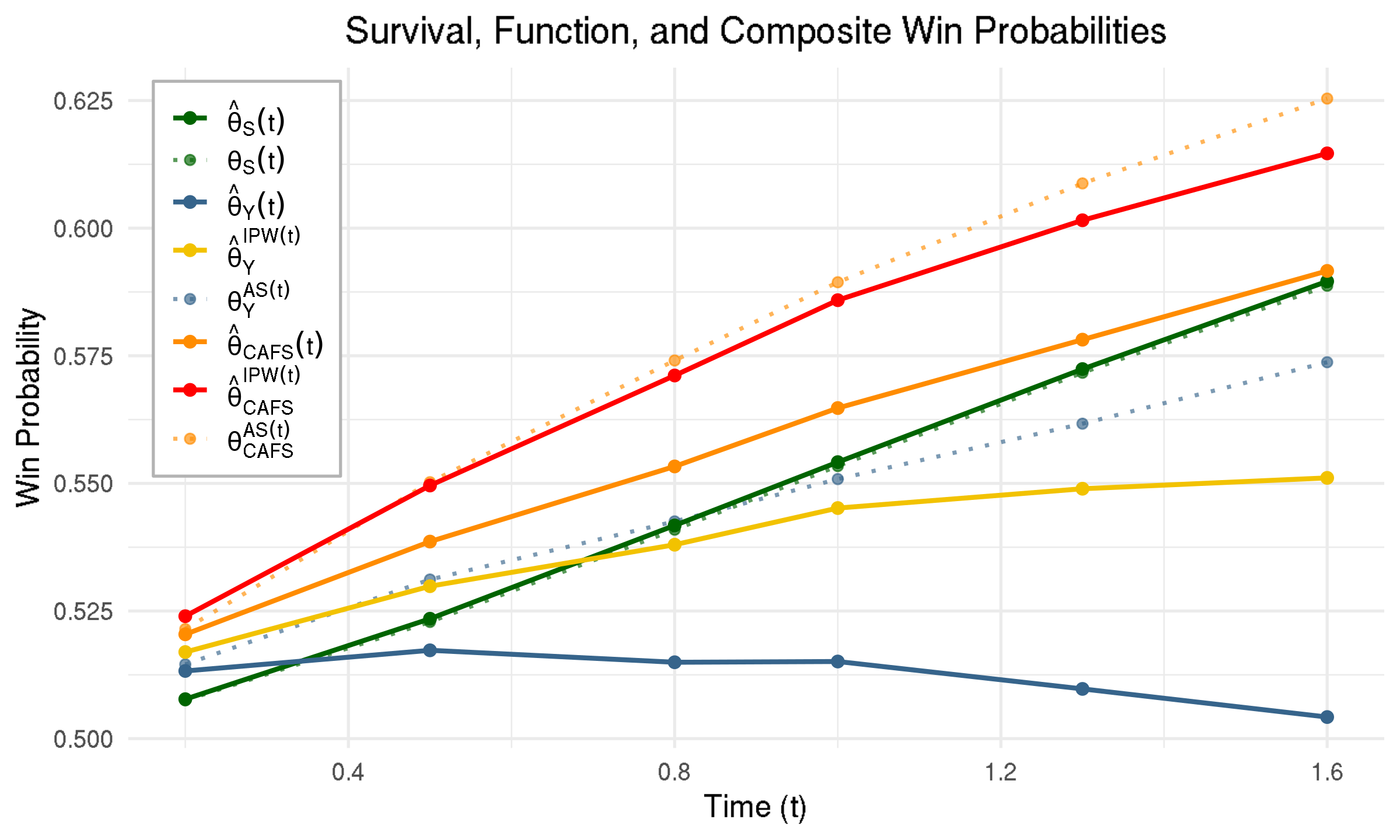}
  \caption{Monte Carlo estimates for Setting~3, when $g_1 = \log(0.5)$ and ${\color{black}B_4} = 2$. Solid lines show observed-data Monte Carlo estimates. Dotted lines show reference values under the counterfactual data-generating model for $\theta_S(t)$, $\theta_Y^{\mathrm{AS}}(t)$, and $\theta_{\mathrm{CAFS}}^{\mathrm{AS}}(t)$. Time $(t)$ is in years.}
  \label{fig:scenario3}
\end{figure}

{\color{black}Figure~\ref{fig:scenario3} makes the mechanism in Setting~3 explicit by displaying the functional-outcome summaries, the composite summaries, and the survival contribution together. $\widehat\theta_Y(t)$ remains only modestly above $0.5$, with a small early increase followed by attenuation over follow-up, even though the data-generating mechanism includes a direct functional benefit. This attenuation should not be interpreted as the longitudinal treatment effect alone because the observed-survivor comparison is restricted to participants who remain alive and observable at each time point and therefore reflects both the functional trajectory and changes in the survivor population over time. The dotted $\theta_Y^{\mathrm{AS}}(t)$ curve lies above $\widehat\theta_Y(t)$, consistent with selection attenuating the observed survivor-based comparison. In contrast, $\widehat\theta_Y^{\mathrm{IPW}}(t)$ is larger than $\widehat\theta_Y(t)$ and rises before flattening at later time points, consistent with IPW attenuating, but not eliminating, selection induced by differential survival under the weighting assumptions. The remaining flattening of the IPW functional summary is compatible with its interpretation as an adjusted while-alive comparison under the weighting assumptions rather than a fixed-population functional effect. Participants with lower baseline functional status or poorer prognosis tend to die earlier, so later comparisons are made in a more selected and comparatively homogeneous survivor population. Meanwhile, $\theta_S(t)$ increases steadily, so the survival contribution becomes increasingly influential in the composite. Consequently, both $\widehat\theta_{\mathrm{CAFS}}(t)$ and $\widehat\theta_{\mathrm{CAFS}}^{\mathrm{IPW}}(t)$ increase over time. At later time points, the upward movement of the composite reflects the growing survival contribution to the composite ranking rather than a strengthening of the observed survivor-based functional comparison.}

\subsection{Discussion of the Three Settings}

{\color{black}Taken together, the three settings clarify how survival-prioritized hierarchical composite endpoints behave when survival and functional outcomes align or conflict, using CAFS as the ALS-specific implementation. In Setting~1, survival benefit alone induces survivor-selection bias in $\widehat\theta_Y(t)$, and $\widehat\theta_Y^{\mathrm{IPW}}(t)$ can attenuate that bias under the weighting assumptions. In Setting~2, functional benefit with worse survival shows that the functional-outcome and composite summaries can move in opposite directions, with the survival-first ranking increasingly dominating $\widehat\theta_{\mathrm{CAFS}}(t)$ and $\widehat\theta_{\mathrm{CAFS}}^{\mathrm{IPW}}(t)$ at later follow-up. In Setting~3, benefit in both domains shows that $\widehat\theta_Y(t)$ can remain attenuated even under a direct functional benefit, while IPW adjustment partially attenuates survivor-selection bias in this comparison; nevertheless, the growing survival contribution is what sustains and strengthens the composite over time.}

{\color{black}The overall message is not that IPW should automatically replace original CAFS. Its role depends on the scientific question, but the decomposition in equation~\eqref{eq:theta_cafs} shows why the survivor-based functional contribution remains important across all three settings because the composite differs from the pure survival comparison. When survival is the primary target and functional outcome is used only to break ties among survivors, $\theta_S(t)$ remains the primary survival comparison. The IPW adjustment $\theta_Y^{\mathrm{IPW}}(t)$ can nevertheless attenuate selection in the functional tie-breaking comparison under the weighting assumptions, but it is not a correction to the survival estimand itself. When functional outcome is the primary target and death is treated as an intercurrent event, IPW adjustment of the survivor-conditional functional comparison is more directly motivated because $\theta_Y(t)$ may be biased by differential survival. A hypothetical strategy could instead define the functional contrast under a no-death or common-survival scenario, but that target would require additional assumptions about how functional outcomes are defined or modeled when death would otherwise preclude measurement \cite{ICH2019E9R1}. Even there, however, the target must be stated carefully. The quantity $\theta_Y^{\mathrm{IPW}}(t)$ is an adjusted survivor-based or while-alive comparison under modeling assumptions. When the scientific target is overall treatment benefit, substituting $\theta_Y^{\mathrm{IPW}}(t)$ into equation~\eqref{eq:theta_cafs} yields an adjusted composite summary that can attenuate, but need not eliminate, selection in the survivor-based functional contribution while retaining the survival-first ranking.}

{\color{black}Thus, both original CAFS and IPW-CAFS remain survival-first composite summaries rather than functional-outcome-specific estimands, although IPW-CAFS targets an adjusted composite summary rather than the original CAFS estimand. More generally, the composite result depends on both the direction and the relative magnitude of survival and functional outcomes. Our simulations show that the survival contribution can become especially influential as follow-up increases and survival differences accumulate. Ultimately, both CAFS summaries remain composite summaries, and their interpretation depends on how survival and functional outcomes jointly contribute to the ranking.}

\subsection{Additional Discussion}

\subsubsection{Remark on the Functional Tie-Breaker and Later Survival}
\label{sec:diagnostic-agreement}

{\color{black}When function serves only to break survival ties, a related question is whether its ordering agrees with later survival. We compare the net functional ordering among pairs alive at $t$ with the additional survival ordering accumulated by a prespecified horizon $s>t$. This comparison is descriptive and should not be interpreted as an individual-level rank-preservation condition or a formal surrogacy criterion.}

{\color{black}Using the notation introduced in Section~2, the CAFS win probability can be written as the survival win probability plus the net functional ordering among pairs alive at $t$, scaled by the probability that both members are alive:}
\begin{align}
\theta_{\mathrm{CAFS}}(t)
&= \theta_S(t)
+ \frac{\Pr(T^{(1)}>t,\,T^{(0)}>t)}{2}
\Big[
\Pr\!\big(Y^{(1)}(t)>Y^{(0)}(t)\mid T^{(1)}>t,\,T^{(0)}>t\big)
\nonumber\\
&\hspace{5.1cm}
-\Pr\!\big(Y^{(1)}(t)<Y^{(0)}(t)\mid T^{(1)}>t,\,T^{(0)}>t\big)
\Big]. \label{eq:theta-decomp}
\end{align}
{\color{black}Here, the survival win probability is}
\begin{equation}
\theta_S(t)= -\int_{0}^{t} S_1(u)\,dS_0(u) + \frac12\,S_1(t)S_0(t). \label{eq:thetaS-diagnostic}
\end{equation}
{\color{black}For a prespecified later horizon $s>t$, the additional survival ordering accumulated between $t$ and $s$ is}
\begin{equation}
\theta_S(s)-\theta_S(t)
= \frac12\!\left[ -\!\int_{t}^{s} S_1(u)\,dS_0(u)
+ \int_{t}^{s} S_0(u)\,dS_1(u) \right]. \label{eq:resid-benefit}
\end{equation}

{\color{black}Exact agreement between the functional ordering at $t$ and the additional survival ordering through $s$ would correspond to}
\begin{equation}
\Delta(s,t)=0, \label{eq:diagnostic-agreement-condition}
\end{equation}
{\color{black}where the diagnostic gap is defined as}
\begin{align}
\Delta(s,t)
&:= \big[\theta_S(s)-\theta_S(t)\big]
-\frac{\Pr(T^{(1)}>t,\,T^{(0)}>t)}{2}
\Big[
\Pr\!\big(Y^{(1)}(t)>Y^{(0)}(t)\mid T^{(1)}>t,\,T^{(0)}>t\big) \nonumber\\
&\hspace{4.8cm}
-\Pr\!\big(Y^{(1)}(t)<Y^{(0)}(t)\mid T^{(1)}>t,\,T^{(0)}>t\big)
\Big]. \label{eq:gap}
\end{align}
{\color{black}The gap $\Delta(s,t)$ measures the difference between the functional ordering at $t$ and the additional survival ordering from $t$ to $s$. These two population-level pairwise summaries can differ even when function is strongly associated with later survival because the functional ordering depends on who remains alive at $t$, whereas the additional survival ordering depends on events accumulated after $t$. In Setting~1 ($B_4=0$), $\Delta(s,t)\neq 0$ indicates this mismatch and can also reflect treatment-induced selection among survivors at $t$. The survivor-based functional comparison should therefore not generally be interpreted as a surrogate or direct proxy for later survival. Supporting figures are provided in the Appendix.}

\subsubsection{Counterfactual Data and Principal-Stratum Estimands}
\label{subsec:counterfactual-ps}

{\color{black}To evaluate principal-stratum and composite win-probability estimands, we generated counterfactual potential outcomes under one control scenario and three treatment scenarios. This construction yields known survival, principal-stratum functional, and composite reference values under a fully specified joint model, allowing the direct treatment parameters for survival and function to vary separately.}
The causal notation and identification assumptions used for these principal-stratum targets are summarized in Appendix~\ref{subsec:principal-stratum-assumptions}.

We considered one control scenario and three treatment scenarios corresponding to (1) survival benefit only, (2) {\color{black}functional benefit only}, and (3) simultaneous survival and functional benefit. In the counterfactual notation below, $a\in\{0,1\}$ denotes a possible value of the treatment assignment $A_i$, with $a=0$ for the shared control setting and $a=1$ for the corresponding scenario-specific treatment setting.

\begin{table}[H]
\centering
\caption{\small Arm-specific parameters by scenario.}
\label{tab:dgp-arms}
\small
\begin{tabularx}{\textwidth}{>{\raggedright\arraybackslash}p{3.6cm}ccX}
\toprule
Scenario & ${\color{black}B_4}$ & $g_1$ & Interpretation \\
\midrule
Control      & $0$     & $0$            & No survival or functional benefit \\
Survival benefit only  & $0$     & $\log(0.5)$   & \textcolor{black}{Direct survival benefit, with no direct functional effect} \\
\textcolor{black}{Functional benefit only}  & $2$    & $0$            & Direct functional benefit, with no direct survival effect \\
\textcolor{black}{Survival and functional benefit}  & $2$    & $\log(0.5)$   & Direct survival and functional benefit \\
\bottomrule
\end{tabularx}
\end{table}

For each participant, we simulated repeated visits at $t\in\{0.25,0.50,\ldots,1.60\}$ and generated potential survival times and functional outcomes under each scenario. {\color{black}We first drew $(B_{0i},B_{1i},B_{2i})^\top$ from $N\{(B_0,B_1,B_2)^\top,\mathbf{\Sigma}\}$ to induce between-participant heterogeneity in baseline function, linear rate of decline, and quadratic curvature. Conditional on these random effects and the arm-specific parameters $(B_4,g_1)$, we generated the functional trajectory and survival time from the joint longitudinal--survival model.}  
This yields, for each participant and each scenario, 
a potential survival time, an at-risk indicator at each visit, and the corresponding functional outcome trajectory. For
the principal-stratum summaries that impose monotonicity, the counterfactual survival construction was implemented to
satisfy the relevant no-harm ordering in the survival-benefit scenarios; if an applied implementation does not enforce this
ordering, the harmed-survivor stratum should be reported rather than omitted.
This counterfactual construction allows direct evaluation of estimands defined on survival-based principal strata and clarifies how observed-survivor comparisons can deviate from the intended causal contrast when treatment affects survival.

At a fixed time $t$, define
\[
D_i(a,t)=I\{T_i(a)>t\}, \qquad a\in\{0,1\},
\]
and
\[
U_i(t)=\bigl(D_i(1,t),D_i(0,t)\bigr).
\]
The 
principal strata are then: 
always-survivors (AS), with $U_i(t) = (1,1)$; 
protected-survivors (PS), with $U_i(t) = (1,0)$; 
harmed-survivors (HS), with $U_i(t) = (0,1)$; 
and never-survivors (NS), with $U_i(t) = (0,0)$. %

Let $h(u,v)=I(u>v)+\tfrac12 I(u=v)$, and let $i$ and $j$ denote independent draws from the treatment and control potential-outcome distributions under the scenario under consideration. The always-survivor functional win probability is a pairwise principal-stratum estimand,
\[
\theta_Y^{\mathrm{AS}}(t)
=
E\!\left[
h\{Y_i(1,t),Y_j(0,t)\}
\mid U_i(t)=U_j(t)=(1,1)
\right].
\]
Thus, $\theta_Y^{\mathrm{AS}}(t)$ compares a treated potential outcome and a control potential outcome for two independent individuals who would both survive beyond $t$ under either treatment.

For protected-survivor and harmed-survivor strata, one of the two treatment-specific functional outcomes at time $t$ is not an ordinary while-alive outcome because the participant would be dead under one treatment condition. In the counterfactual simulation below, these quantities are therefore interpreted as latent or hypothetical functional-trajectory contrasts generated by the joint model, rather than as directly observable while-alive functional outcomes:
\[
\theta_Y^{\mathrm{PS}}(t)
=
E\!\left[
h\{Y_i(1,t),Y_j(0,t)\}
\mid U_i(t)=U_j(t)=(1,0)
\right],
\]
and, when present, 
\[
\theta_Y^{\mathrm{HS}}(t)
=
E\!\left[
h\{Y_i(1,t),Y_j(0,t)\}
\mid U_i(t)=U_j(t)=(0,1)
\right].
\]

When the monotonicity assumption $T_i(1) \ge T_i(0)$ is imposed, corresponding to no treatment-induced survival harm, the harmed-survivor stratum is empty. In that case $\theta_Y^{\mathrm{HS}}(t)$ is undefined and omitted from estimation. When monotonicity is not plausible or not enforced by the simulation design, $\theta_Y^{\mathrm{HS}}(t)$ should be retained as a separate stratum-specific quantity. The never-survivor stratum contributes no functional-outcome information at time $t$.

{\color{black}We summarize win-probability estimates at $t=1.6$ years for the always-survivor principal-stratum estimand $\theta_Y^{\mathrm{AS}}(t)$, the protected-survivor quantity $\theta_Y^{\mathrm{PS}}(t)$, and the observed-survivor benchmark $\theta_Y^{\mathrm{OS}}(t)$. In this counterfactual simulation, $\theta_Y^{\mathrm{OS}}(t)$ is the large-sample analogue of $\theta_Y(t)$ obtained by comparing participants who are alive under their realized treatment condition.} To quantify its discrepancy from the always-survivor target, we report
\[
\mathrm{Gap}(t) = \theta_Y^{\mathrm{AS}}(t) - \theta_Y^{\mathrm{OS}}(t),
\]
together 
with the estimated numbers of always-survivors ($N_{\mathrm{AS}}$) and protected-survivors ($N_{\mathrm{PS}}$).

\begin{table}[H]
\centering
\caption{\small Monte Carlo estimates of win probabilities at $t=1.6$ years ($N=10{,}000$ per scenario).}
\label{tab:mc-results-counterfactual}
\small
\begin{tabularx}{\textwidth}{>{\raggedright\arraybackslash}Xcccccc}
\toprule
Scenario & $\theta_Y^{\mathrm{AS}}$ & $\theta_Y^{\mathrm{PS}}$ & $\theta_Y^{\mathrm{OS}}$ & $\mathrm{Gap}(t)$ & $N_{\mathrm{AS}}$ & $N_{\mathrm{PS}}$ \\
\midrule
Control & 0.500 & -- & 0.500 & 0.000 & 6379 & 0 \\
Survival benefit only & 0.500 & 0.500 & 0.453 & 0.047 & 6379 & 1255 \\
\textcolor{black}{Functional benefit only} & 0.575 & 0.589 & 0.554 & 0.021 & 6379 & 509 \\
\textcolor{black}{Survival and functional benefit} & 0.575 & 0.583 & 0.511 & 0.064 & 6379 & 1656 \\
\bottomrule
\end{tabularx}
\end{table}

Table~\ref{tab:mc-results-counterfactual} makes the estimand gap explicit at the same endpoint used for the IPW comparisons. In the survival-benefit-only scenario, $\theta_Y^{\mathrm{AS}}(t) \approx 0.5$ under ${\color{black}B_4}=0$, whereas $\theta_Y^{\mathrm{OS}}(t)$ falls to 0.453. In the analogous observed-data setting in Table~\ref{tab:mc-results}, $\widehat\theta_Y^{\mathrm{IPW}}(t)=0.493$ at $t=1.6$, closer to the always-survivor value but still distinct from it. {\color{black}The observed-survivor benchmark also attenuates the always-survivor contrast in the functional-benefit-only scenario, from 0.575 to 0.554, and in the combined-benefit scenario, from 0.575 to 0.511.} These examples show that IPW adjustment may move the comparison toward the always-survivor contrast without changing its interpretation as an adjusted while-alive target.
This distinction is also important substantively. The always-survivor estimand targets only the subpopulation who would survive under either treatment, so it need not be the preferred target for the primary trial question. When the clinical aim is to describe treatment benefit for all randomized participants, an estimand defined only within the always-survivor stratum may be too narrow even if it is useful for isolating function from differential survival.

\subsubsection{Supporting Illustration: Multiply Robust Estimation of the Always-Survivor Principal-Stratum Win Probability}
\label{subsec:mr-estimator}

Because the always-survivor stratum is latent, estimation of $\theta_Y^{\mathrm{AS}}(t)$ requires additional modeling. {\color{black}We include multiply robust estimation as a supporting analysis when the scientific question concerns function among always-survivors rather than overall benefit in the randomized population.}

Operationally, we fit (i) a survival or principal-score model to estimate weights for membership in the always-survivor stratum and (ii) a working model for the functional win probability. The resulting multiply robust estimator combines a model-based plug-in component with an inverse-probability-weighted 
correction term. Full 
derivations and influence-function details are provided in the Appendix and in \cite{jiang2022multiply,chen2024principal}.
\begin{figure}[H]
\centering
\includegraphics[width=0.95\textwidth]{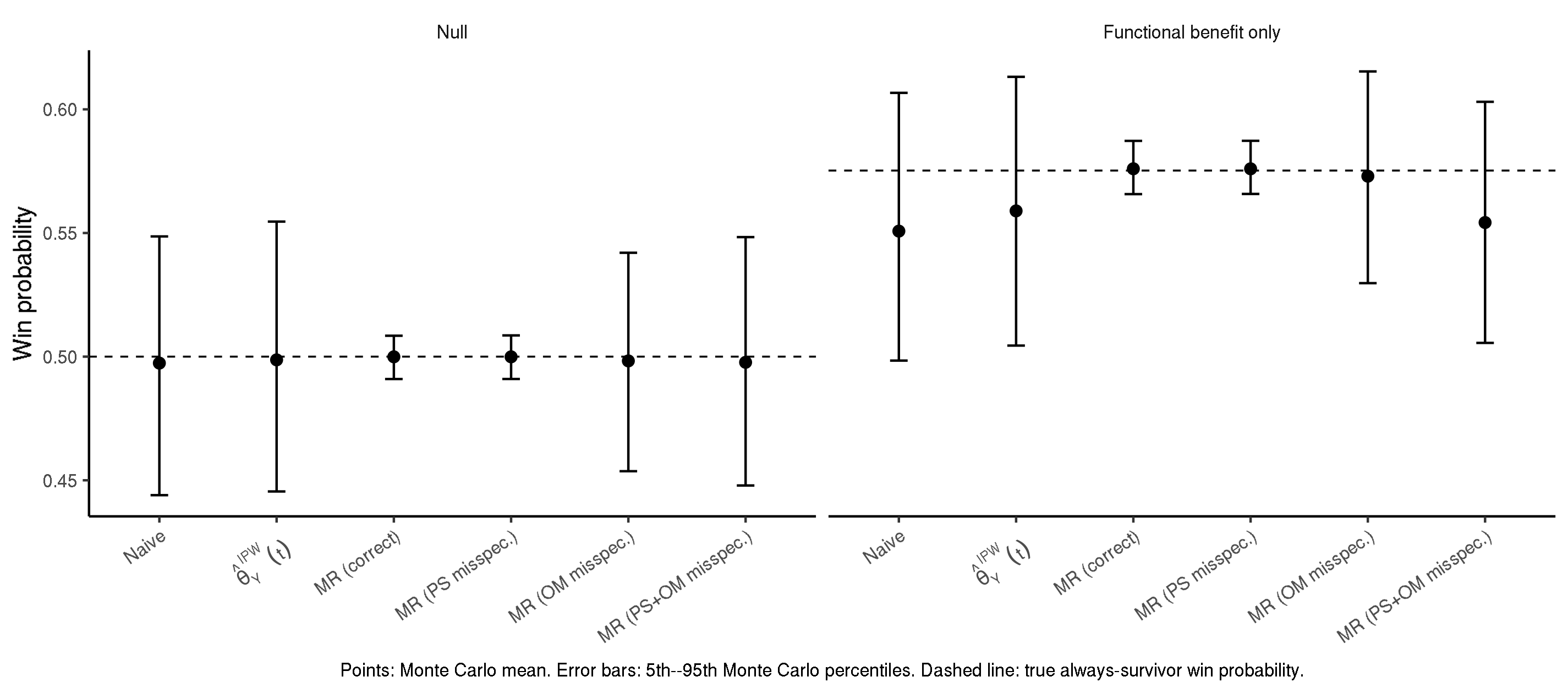}
\captionsetup{font=footnotesize}
\caption{Supporting Monte Carlo illustration for the always-survivor win probability at $t=1.6$ years. The naive observed-survivor estimator and $\widehat{\theta}_Y^{\mathrm{IPW}}(t)$ are included as references; the multiply robust estimators target $\theta_Y^{\mathrm{AS}}(t)$ under correct specification and under single or double nuisance-model misspecification. Points show Monte Carlo means and bars denote the 5th--95th percentiles; dashed lines indicate the true always-survivor win probability.}
\label{fig:wp-mr-sim}
\end{figure}
Figure~\ref{fig:wp-mr-sim} summarizes Monte Carlo results based on 1,000 replications with $N=500$ at $t=1.6$ years. In the null scenario, the true always-survivor win probability is 0.500, and the Monte Carlo means are centered near 0.500. {\color{black}In the functional-benefit-only scenario, the true always-survivor win probability is 0.575. The naive observed-survivor estimator and $\widehat{\theta}_Y^{\mathrm{IPW}}(t)$ are shown as references and have lower means of 0.551 and 0.559, respectively. The multiply robust estimators for which at least one nuisance model is correctly specified are centered near the true always-survivor value, whereas the estimator with both nuisance models misspecified is lower.} We include these results to show that, once the always-survivor estimand is chosen as the scientific target, multiply robust estimation can serve as a supporting analysis under the stated principal-stratum identification assumptions.

\section{Conclusion}
{\color{black}We examined hierarchical composite endpoints in trials with both survival and functional outcomes, focusing on pairwise comparisons of the Finkelstein--Schoenfeld type and using CAFS in ALS as a motivating implementation. The central issue is estimand clarification. Depending on the scientific question, interest may center on survival with function used to break ties among survivors, function with death treated as an intercurrent event, or overall treatment benefit combining survival and function. Survival-prioritized hierarchical composites such as CAFS target the third question. Decomposing the composite helps investigators understand the relative survival and functional contributions while distinguishing the overall composite target from secondary survival- or function-focused questions.}

{\color{black}The three simulation settings make this point concrete. In Setting~1, survival benefit alone induces survivor-selection bias in the observed survivor-based functional comparison. In Setting~2, functional benefit with survival harm shows that the functional and composite summaries can move in opposite directions, with the survival-first ranking increasingly dominating the composite at later follow-up. In Setting~3, benefit in both domains shows how a growing survival advantage can sustain and strengthen the composite even when the observed survivor-based functional comparison attenuates and IPW only partially attenuates survivor-selection bias. Expressing the composite win probability as the sum of survival and survivor-based functional contributions explains these patterns.}

{\color{black}Within this framework, IPW-adjusted composite summaries should be viewed and reported as adjusted hierarchical composite summaries rather than as estimates of the original CAFS estimand. Under credible weighting assumptions, they can attenuate selection induced by measured predictors in the survivor-conditional functional comparison before that comparison is combined with survival. They do not turn a hierarchical composite endpoint into a function-only measure. When overall treatment benefit under the CAFS ranking is the prespecified target, the original CAFS estimate should remain the primary analysis and should be reported with its survival and functional contributions. Hypothetical and principal-stratum strategies can define alternative function-focused targets, but they require additional modeling or identification assumptions. The principal-stratum and multiply robust analyses are included as supporting tools for the always-survivor target when the scientific aim is to isolate function from differential survival. Future work will extend this framework to settings with missed visits, dropout, censoring, and other forms of missing longitudinal data and will develop hypothesis-testing and inferential procedures for hierarchical composite endpoints.}









\printbibliography[title={REFERENCES }, heading=bibintoc]

\appendix
\section{Supporting Methods and Figures}

\subsection{Supporting Figures for the Functional Tie-Breaker and Later Survival Remark}

{\color{black}This appendix provides supporting figures for the population-level comparison developed in Section~\ref{sec:diagnostic-agreement}. The decomposition, residual survival ordering, and gap are defined in equations~\eqref{eq:theta-decomp}--\eqref{eq:gap}.}

\begin{figure}[H]
  \centering
  {\color{black}%
  \resizebox{0.6\linewidth}{!}{%
    \setlength{\unitlength}{1bp}%
    \begin{picture}(1353,1163)
      \put(0,0){\includegraphics[width=1353bp,height=1163bp]{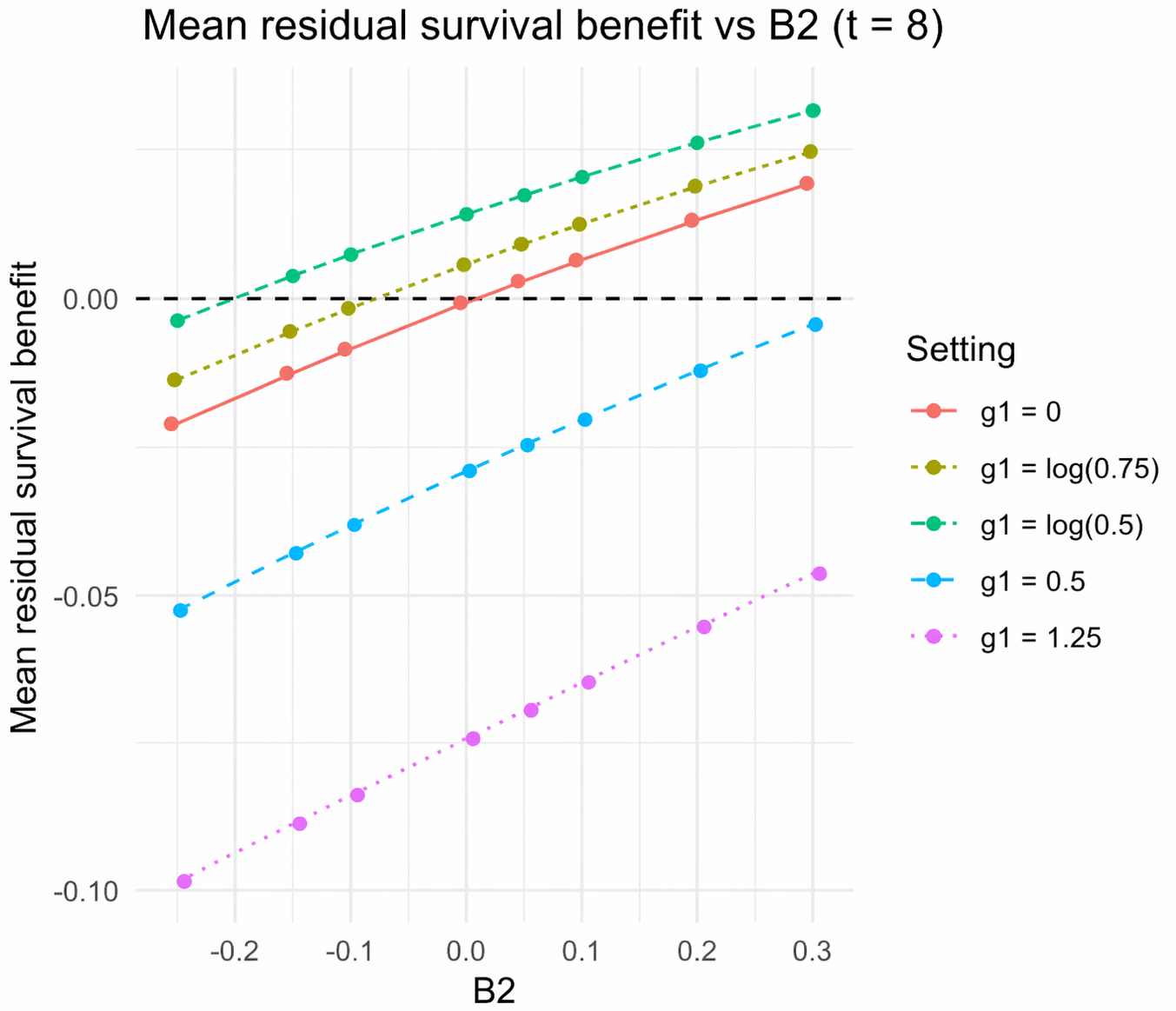}}
      \put(95,1082){{\color{white}\rule{1000bp}{81bp}}}
      \put(95,1097){\makebox(1000,55)[c]{\sffamily\fontsize{24.88}{28}\selectfont Mean residual survival benefit vs $B_4$ $(t=8)$}}
      \put(500,0){{\color{white}\rule{250bp}{48bp}}}
      \put(500,2){\makebox(250,40)[c]{\sffamily\fontsize{24.88}{28}\selectfont $B_4$}}
    \end{picture}%
  }}
  \caption{Auxiliary diagnostic parameter sweep of the residual survival ordering from $t=8$ months to $s=24$ months.}
  \label{fig:resid-benefit-null}
\end{figure}
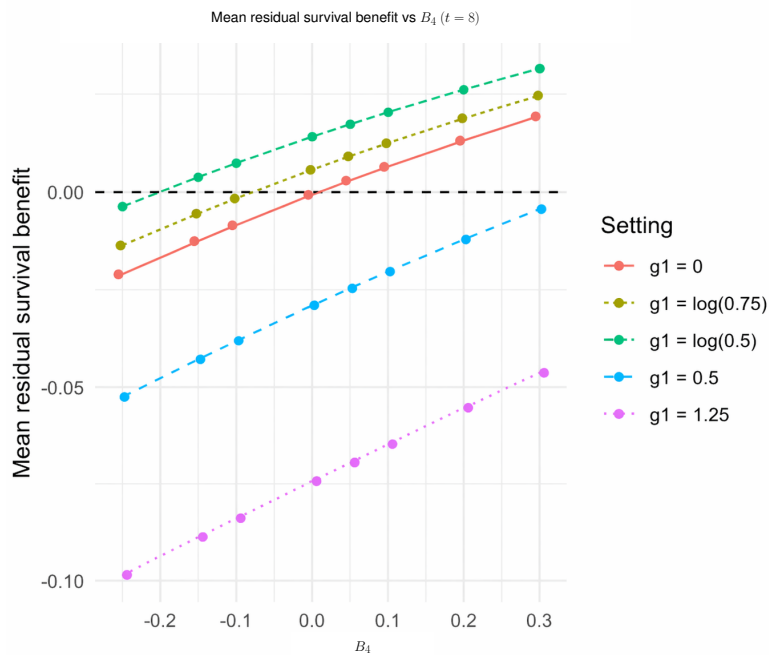

Figure~\ref{fig:resid-benefit-null} is an auxiliary diagnostic sweep under the same joint-model structure, rather than an additional primary simulation setting. We varied the functional treatment effect ${\color{black}B_4}$ while fixing the direct survival effect $g_1$ and evaluated when the mean residual survival ordering in equation~\eqref{eq:resid-benefit} departs from $0$ across 500 Monte Carlo replications, with $t=8$ months and $s=24$ months. When $g_1=0$, the residual ordering remains near zero for small ${\color{black}B_4}$ and becomes positive once improvement in the functional trajectory generates a measurable survival advantage. A direct survival effect shifts the curve upward or downward depending on $g_1$, as expected from equation~\eqref{eq:thetaS-diagnostic}.

To further illustrate the mechanism underlying the decomposition in equation~\eqref{eq:theta-decomp}, for each pair $(g_1,{\color{black}B_4})$ we computed the mean survival contribution and the mean survivor-based functional contribution at $t=8$ months.

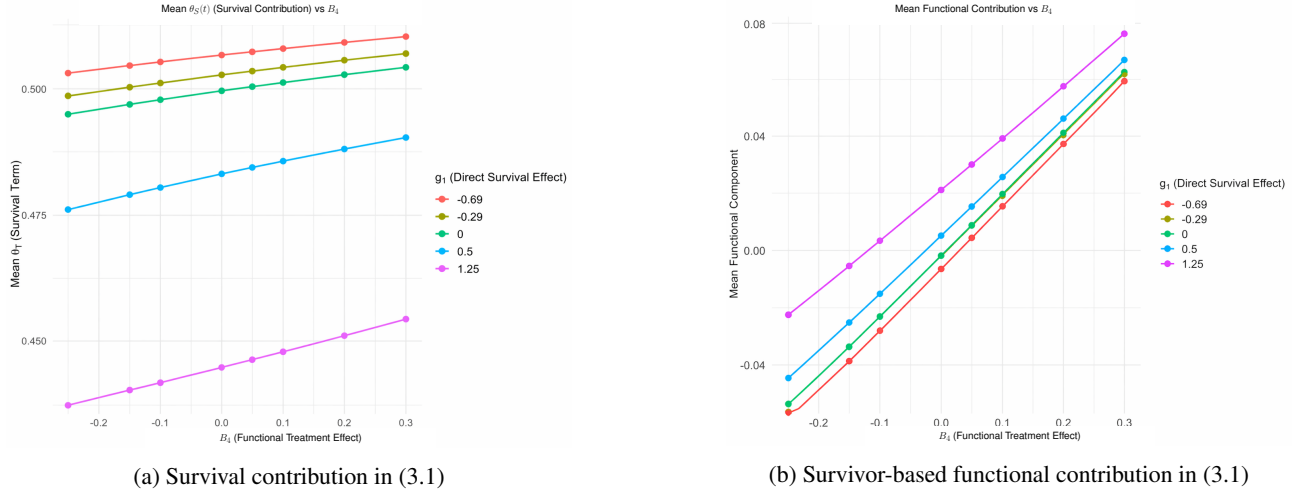
\begin{figure}[H]
\centering

\begin{subfigure}[t]{0.44\textwidth}
    \centering
    {\color{black}%
    \resizebox{\textwidth}{!}{%
      \setlength{\unitlength}{1bp}%
      \begin{picture}(1402,1122)
        \put(0,0){\includegraphics[width=1402bp,height=1122bp]{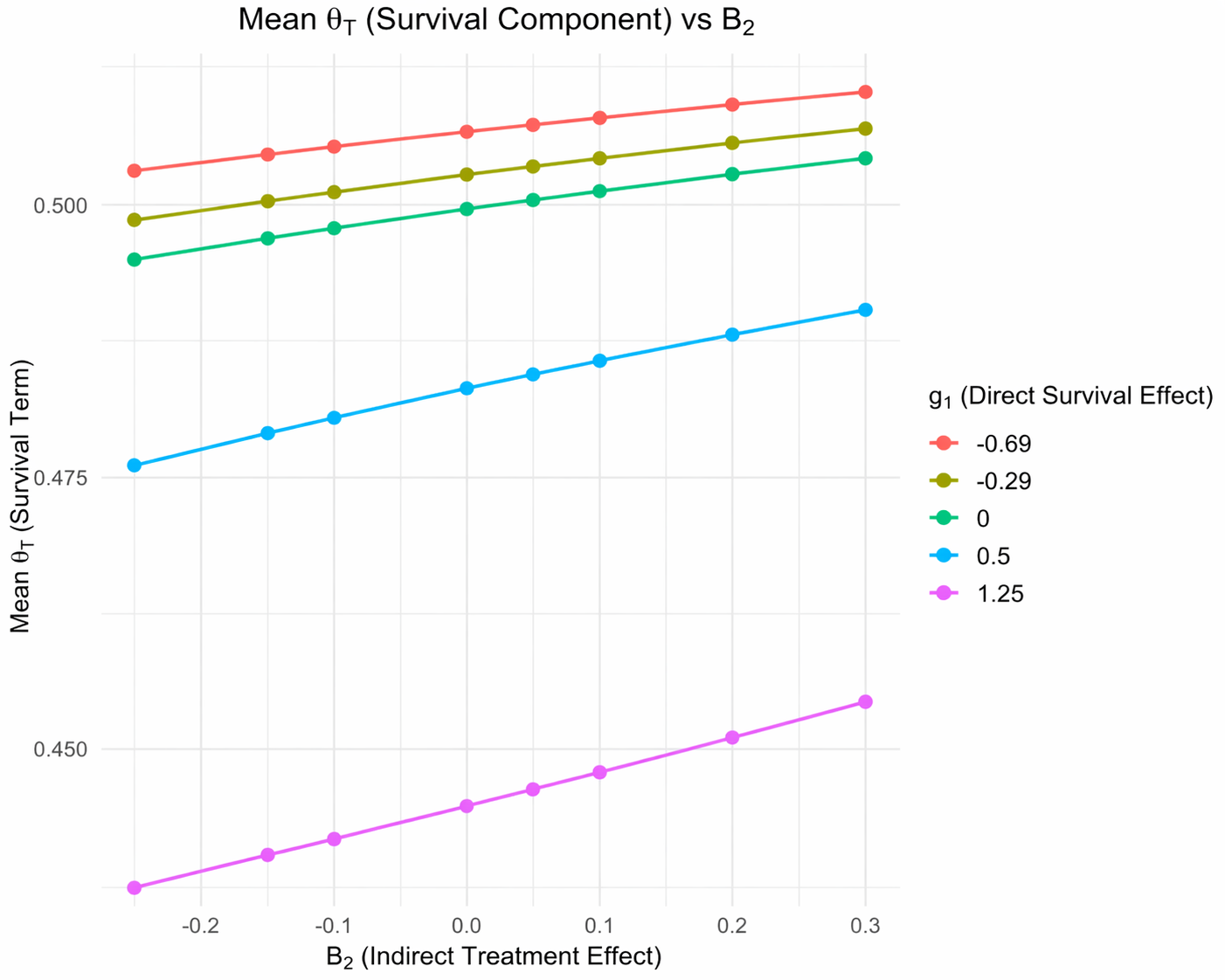}}
        \put(180,1048){{\color{white}\rule{850bp}{74bp}}}
        \put(180,1063){\makebox(850,48)[c]{\sffamily\fontsize{24.88}{28}\selectfont Mean $\theta_S(t)$ (Survival Contribution) vs $B_4$}}
        \put(345,0){{\color{white}\rule{720bp}{48bp}}}
        \put(345,2){\makebox(720,40)[c]{\sffamily\fontsize{24.88}{28}\selectfont $B_4$ (Functional Treatment Effect)}}
      \end{picture}%
    }}
    \caption{Survival contribution in \eqref{eq:theta-decomp}}
    \label{fig:thetaS_vs_B4}
\end{subfigure}
\hfill
\begin{subfigure}[t]{0.44\textwidth}
    \centering
    {\color{black}%
    \resizebox{\textwidth}{!}{%
      \setlength{\unitlength}{1bp}%
      \begin{picture}(1405,1119)
        \put(0,0){\includegraphics[width=1405bp,height=1119bp]{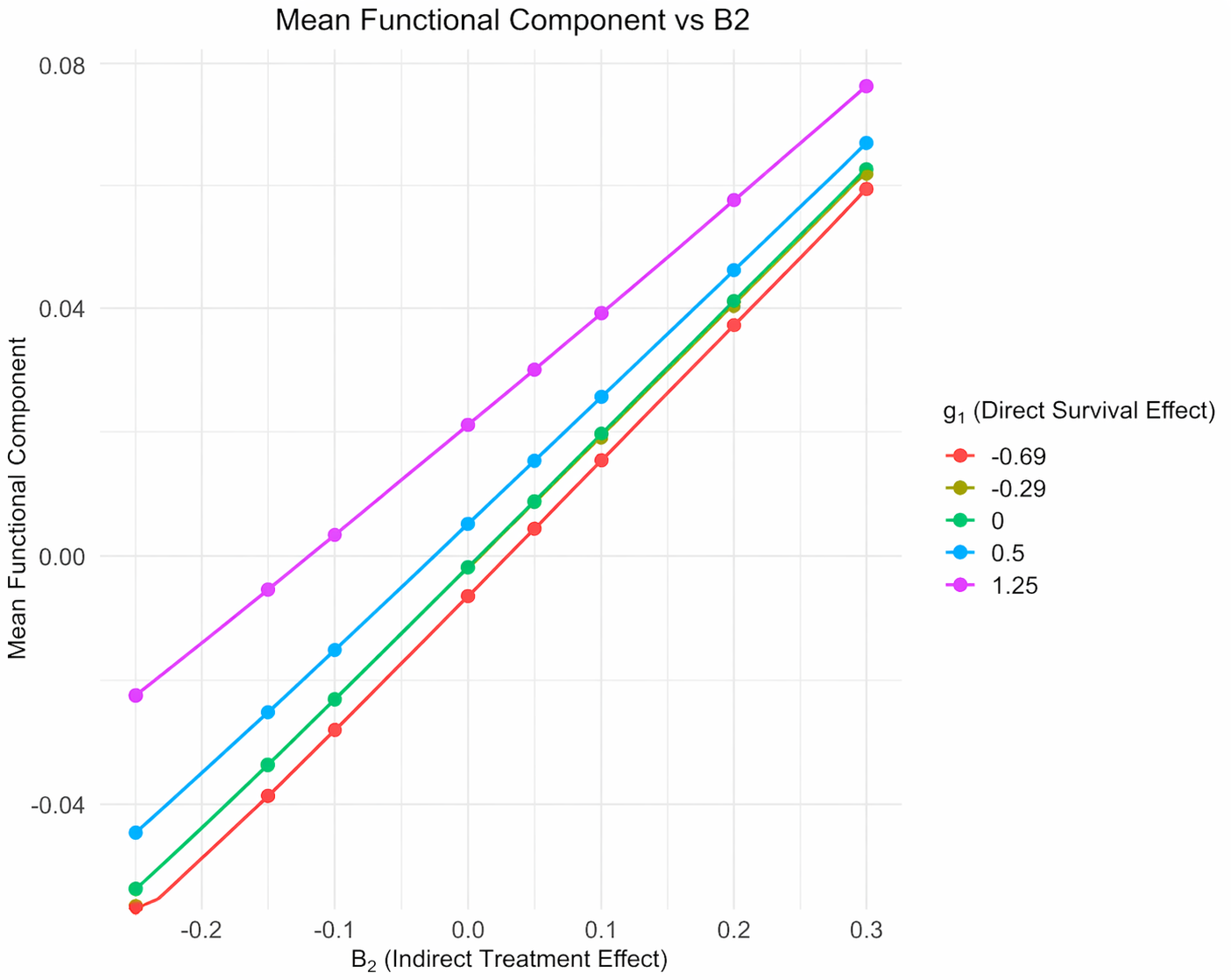}}
        \put(215,1045){{\color{white}\rule{800bp}{74bp}}}
        \put(215,1060){\makebox(800,48)[c]{\sffamily\fontsize{24.88}{28}\selectfont Mean Functional Contribution vs $B_4$}}
        \put(345,0){{\color{white}\rule{720bp}{48bp}}}
        \put(345,2){\makebox(720,40)[c]{\sffamily\fontsize{24.88}{28}\selectfont $B_4$ (Functional Treatment Effect)}}
      \end{picture}%
    }}
    \caption{Survivor-based functional contribution in \eqref{eq:theta-decomp}}
    \label{fig:functional_vs_B4}
\end{subfigure}

\caption{Auxiliary diagnostic decomposition of $\theta_{\mathrm{CAFS}}(t)$ into survival and survivor-based functional contributions at $t=8$ months.}
\label{fig:theta_decomp}
\end{figure}

Figure~\ref{fig:thetaS_vs_B4} shows that the survival contribution increases with ${\color{black}B_4}$ across the displayed settings and shifts upward or downward with $g_1$, reflecting the accumulated survival advantage or harm. Figure~\ref{fig:functional_vs_B4} shows that the survivor-based functional contribution also varies systematically with both ${\color{black}B_4}$ and $g_1$. Importantly, the curves in Figure~\ref{fig:functional_vs_B4} do not overlap across different values of $g_1$, even though the functional-outcome model is unchanged across those $g_1$ values. {\color{black}This separation reflects selection among observed survivors and helps explain why the functional and subsequent survival orderings may disagree.}

\subsection{Principal-Stratum Notation and Assumptions}
\label{subsec:principal-stratum-assumptions}

The quantities defined in the main text are based on observed-arm comparisons between treated and control participants. If the goal is instead to define a function-specific causal effect within a principal stratum, then additional causal notation and assumptions are required.

For each participant $i$ and treatment level $a\in\{0,1\}$, let $T_i(a)$ denote the potential survival time under treatment $a$, and let $Y_i(a,t)$ denote the corresponding potential functional outcome at time $t$, when defined. For always-survivor estimands, both treatment-specific functional outcomes at time $t$ are meaningful while-alive outcomes; for protected- or harmed-survivor contrasts, any functional outcome under the treatment condition in which the participant would not survive to $t$ must be interpreted as a latent or hypothetical trajectory value supplied by an additional model. Since the functional outcome at time $t$ is only meaningful for participants who would survive beyond $t$, define
\[
D_i(a,t)=I\{T_i(a)>t\}, \qquad a\in\{0,1\},
\]
and let
\[
U_i(t)=\bigl(D_i(1,t),D_i(0,t)\bigr)
\]
denote the principal stratum at time $t$, with the always-survivor stratum corresponding to $U_i(t)=(1,1)$.

Write $q_a(X_i,t)=P\{D_i(a,t)=1\mid X_i\}$ for $a\in\{0,1\}$ and $\pi(X_i)=P(A_i=1\mid X_i)$.

In addition to consistency, no interference, randomized treatment assignment, and the conditions required for the weighting estimator, identification of the always-survivor win probability requires the following assumptions. Let $i$ and $j$ denote independent treated and control draws, respectively. A pairwise mean version of principal ignorability is
\[
\begin{aligned}
\textbf{(A1) Pairwise mean principal ignorability:}\quad
&E\!\left[h\{Y_i(1,t),Y_j(0,t)\}\mid U_i(t)=U_j(t)=(1,1),X_i,X_j\right]\\
&=E\!\left[h\{Y_i(t),Y_j(t)\}\mid A_i=1,R_i(t)=1,A_j=0,R_j(t)=1,X_i,X_j\right].
\end{aligned}
\]
\[
\textbf{(A2) Monotonicity: } \quad T_i(1)\ge T_i(0)\ \text{almost surely},
\]
{\color{black}The global survival-time condition in (A2) implies $D_i(1,t)\ge D_i(0,t)$ almost surely at every fixed horizon $t$. If identification is sought only at a prespecified horizon, the weaker fixed-horizon condition $D_i(1,t)\ge D_i(0,t)$ may instead be imposed directly. This fixed-horizon condition does not imply the global ordering $T_i(1)\ge T_i(0)$.} Monotonicity states that treatment cannot cause a participant who would have survived to time $t$ under control to die before time $t$ under treatment. Thus, any participant who survives under control is also an always-survivor, and the harmed-survivor stratum $U_i(t)=(0,1)$ is ruled out. This assumption is useful because it simplifies identification of the always-survivor stratum: control-arm survivors provide direct information about participants who would survive under both treatment assignments. However, monotonicity is a substantive clinical assumption, not a statistical consequence of randomization or of the principal-stratum framework. It should therefore be viewed as target- and scenario-specific rather than automatic. If treatment might harm survival, as in settings designed to study survival harm with functional benefit, the harmed-survivor stratum is part of the estimand structure and should not be suppressed.
\[
\textbf{(A3) Positivity: } \quad
\varepsilon<\pi(X_i)<1-\varepsilon,\qquad
q_0(X_i,t)>\varepsilon,\qquad q_1(X_i,t)>\varepsilon
\]
for some $\varepsilon>0$ and all $X_i$ in the relevant covariate support.

Under (A1)--(A3), contrasts within the always-survivor stratum may be identified, and corresponding weighted estimators may be consistent under their stated model conditions. These assumptions are strong and generally not testable. In particular, when $P(U_i(t)=(1,1)\mid X_i)$ is small in some regions of the covariate space, the resulting weights may become unstable, leading to increased variance and possible finite-sample bias.

\subsection{Detailed Construction of the Multiply Robust Estimator}
For completeness, we record the construction used to implement the practical estimator in Section~\ref{subsec:mr-estimator}.
To remain consistent with the notation in the Methods section, we write potential survival status at time $t$ as
\[
D_i(a,t) = I\{T_i(a) > t\}, \qquad \text{for } a \in \{0,1\},
\]
and principal-stratum membership as $U_i(t) = (D_i(1,t),D_i(0,t))$. 
 
{\color{black}Under fixed-horizon monotonicity at time $t$}, $D_i(1,t) \geq D_i(0,t)$ almost surely, so the always-survivor stratum satisfies 
\[
U_i(t) = (1,1) \text{ if and only if } D_i(0,t) = 1.
\]

Recall that
\[
q_a(X_i,t)=P(D_i(a,t)=1\mid X_i), \qquad a\in\{0,1\}.
\]
Then
\[
P(U_i(t)=(1,1)\mid X_i)=q_0(X_i,t).
\]
Among treated participants who survive beyond time $t$, that is, among participants with $A_i=1$ and $R_i(t)=1$, where
\[
R_i(t)=I\{T_i>t\},
\]
monotonicity implies
\[
P(U_i(t)=(1,1)\mid A_i=1,R_i(t)=1,X_i)
=
\frac{q_0(X_i,t)}{q_1(X_i,t)}.
\]
Among control participants with $A_i=0$ and $R_i(t)=1$, membership in the always-survivor stratum occurs with probability one.

Let 

\[
p_{\mathrm{AS}}(t)=P(U_i(t)=(1,1))=P(D_i(0,t)=1),
\]
where the equality follows from monotonicity. 

Following \cite{jiang2022multiply}, estimate $p_{\mathrm{AS}}(t)$ by 
\[
\widehat{p}_{\mathrm{AS}}(t)
=
\frac{1}{n}\sum_{i=1}^n
\left[
I\{A_i=0\}\frac{R_i(t)-\widehat{q}_0(X_i,t)}{1-\pi(X_i)}
+
\widehat{q}_0(X_i,t)
\right],
\]
where $\pi(X_i)=P(A_i=1\mid X_i)$ denotes the treatment propensity score.

Define the normalized always-survivor weights
\[
u_i(t)=\frac{\widehat{q}_0(X_i,t)}{\widehat{p}_{\mathrm{AS}}(t)},
\] 
which approximately reweights the full covariate distribution toward that of the always-survivor stratum.

For the pairwise comparison kernel, let 
\[
h(u,v)=I(u>v)+\tfrac12 I\{u=v\}.
\] 
Define the observed-pair outcome model
\[
m_t(X_i,X_j)
=
E\!\left[h\!\left(Y_i(t),Y_j(t)\right)\mid
A_i=1,R_i(t)=1,A_j=0,R_j(t)=1,X_i,X_j\right].
\]
Under (A1), this observed-pair conditional mean identifies the corresponding conditional win kernel for two always-survivors.

Under a Normal working model, 
\[
m_t(X_i,X_j)
\approx
\Phi\!\left(
\frac{\mu_1(X_i,t)-\mu_0(X_j,t)}
{\sqrt{\sigma_1^2(t)+\sigma_0^2(t)}}
\right).
\]

For clarity, we first define a principal-score/outcome-model plug-in estimator, denoted ``ps--om'' because it combines the principal-score model with the outcome model:
\[
\widehat{\theta}_{\text{ps--om}}(t)
=
\frac{\sum_{i,j} u_i(t) u_j(t)\, m_t(X_i,X_j)}
{\sum_{i,j} u_i(t) u_j(t)}.
\] 
This quantity is useful as a benchmark but is not the plug-in component used in the final plug-in-plus-residual estimator below. Under monotonicity, control participants who survive beyond time $t$ form a direct sample from the always-survivor stratum. This motivates the treatment-principal/outcome-model plug-in component, denoted ``tp--om'':
\[
\widehat{\theta}_{\text{tp--om}}(t)
=
\frac{\sum_{i,j} \tilde w_i(t) \tilde w_j(t)\, m_t(X_i,X_j)}
{\sum_{i,j} \tilde w_i(t) \tilde w_j(t)},
\qquad
\tilde w_i(t)=I\{A_i=0,R_i(t)=1\}\{1-\pi(X_i)\}^{-1}.
\]

Next, define the residual-correction weights 
\[
w_i^{(1)}(t)
=
\frac{\widehat{q}_0(X_i,t)}{\widehat{p}_{\mathrm{AS}}(t)}
\cdot
\frac{1}{\widehat{q}_1(X_i,t)}
\cdot
\frac{1}{\pi(X_i)},
\quad
w_j^{(0)}(t)
=
\frac{1}{\widehat{p}_{\mathrm{AS}}(t)}
\cdot
\frac{1}{1-\pi(X_j)}.
\]
Let
\[
\mathcal{T}(t)=\{i:A_i=1,\ R_i(t)=1\},
\qquad
\mathcal{C}(t)=\{j:A_j=0,\ R_j(t)=1\}
\]
denote the treated and control participants who survive beyond time $t$. Then the treatment-principal/principal-score residual-correction term, denoted ``tp--ps'', is
\[
\widehat{\theta}_{\text{tp--ps}}(t)
=
\frac{
\sum_{i\in\mathcal{T}(t)}
\sum_{j\in\mathcal{C}(t)}
w_i^{(1)}(t)w_j^{(0)}(t)
\left[
h\!\left(Y_i(t),Y_j(t)\right) - m_t(X_i,X_j)
\right]
}{
\sum_{i\in\mathcal{T}(t)}
\sum_{j\in\mathcal{C}(t)}
w_i^{(1)}(t)w_j^{(0)}(t)
}.
\] 

Hence the multiply robust estimator of the always-survivor win probability can be written as 
\[
\widehat{\theta}_Y^{\mathrm{AS,MR}}(t)
=
\widehat{\theta}_{\mathrm{tp\text{-}om}}(t)
+
\widehat{\theta}_{\mathrm{tp\text{-}ps}}(t).
\]
This is the plug-in-plus-residual representation referred to in Section~\ref{subsec:mr-estimator}; the earlier ps--om expression is included only to clarify the notation and to provide a comparator.

By Theorem~2 of Chen and Li \cite{chen2024principal}, this specialization is multiply robust under the assumptions stated above. The treatment block depends on $\pi(\cdot)$, the principal-score block depends on $\{q_0(\cdot,t),q_1(\cdot,t),p_{\mathrm{AS}}(t)\}$, and the outcome block depends on $m_t(\cdot,\cdot)$. If any two of these three nuisance blocks are correctly specified, the first-order remainder vanishes and the estimator is consistent for $\theta_Y^{\mathrm{AS}}(t)$. In a randomized trial with known treatment probability, the treatment block is fixed by design, so the practical robustness is mainly with respect to the principal-score and outcome-model components.

\end{document}